\documentstyle[editedvolume,numreferences]{crckapb} 

\input epsfig

\newcommand{\be}{\begin{equation}}
\newcommand{\ee}{\end{equation}}
\newcommand{\bea}{\begin{eqnarray}}
\newcommand{\eea}{\end{eqnarray}}
\newcommand{\nn}{\nonumber \\}

\def\bfO{\mbox{\boldmath $\Omega$}}
\def\bfs{\mbox{\boldmath $\sigma$}}
\def\bfn{\mbox{\boldmath $\nabla$}}
\def\tr{{\rm tr}\,}

\font\mybb=msbm10 at 10pt
\def\bb#1{\hbox{\mybb#1}}

\def\bR {\bb{R}}
\def\bC {\bb{C}}
\def\bE {\bb{E}}

\def\bI {\bb{I}}

\begin{opening}
\title{BRANE THEORY SOLITONS}
\subtitle{Carg{\`e}se Lectures 1999}
\author{P.K. TOWNSEND}
\institute{CENTRE FOR MATHEMATICAL SCIENCES,\\
           UNIVERSITY OF CAMBRIDGE,\\
           WILBERFORCE ROAD, \\
           CAMBRIDGE, U.K.}
\end{opening}
\runningtitle{BRANE THEORY SOLITONS}

\begin{document}
\begin{abstract}
Field theories that describe {\sl small} fluctuations of branes are limits of
`brane theories' that describe {\sl large} fluctuations. In particular,
supersymmetric sigma-models arise in this way. These lectures discuss the
soliton solutions of the associated `brane theories' and their relation to
calibrations. 
\end{abstract}

\section{Preamble}

The last five years or so have seen some exciting developments in high
energy/gravitational physics, with branes as the common feature. In
particular, branes have revolutionized our ideas about quantum field theory, 
both on the technical level, by giving us new and powerful methods that allow us
to go beyond perturbation theory, and on the conceptual level, by providing us
with a new insight into its nature: it now seems likely that all consistent
quantum field theories can be viewed as effective descriptions of low-energy
fluctuations of branes. 

The small fluctuations of a single brane are governed by a free field theory.
Going beyond small fluctuations, but still on a single brane, introduces
interactions but these are of higher-derivative type, and hence associated with 
a characteristic length scale $L$. The 11-dimensional supermembrane provides a
simple example, with $L$ determined by the membrane tension.
Interactions of conventional field theory type arise from {\sl inter-brane}
interactions. These will dominate if the branes are separated by distances much
less than $L$ but if $L$ also sets the scale for the brane `core', as will
typically be the case, then the inter-brane dynamics cannot be separated from
the unknown core dynamics. Exceptions to this state of affairs can arise only
when there is a separate length scale $L_c$ determined by the size of the core,
with $L_c\ll L$, and this implies the existence of a small dimensionless
constant $g=L_c/L \ll 1$. This scenario is realized by the D-branes of
superstring theory, with
$L=l_s$ the string length (set by the string tension) and $g^{p+1}=g_s$, for a
Dp-brane, with $g_s$ the string coupling constant (which must be small for
superstring theory to be a valid approximation); in this case the inter-brane
interactions between parallel D-branes separated by distances $l$ with $g_s
l_s\ll l\ll l_s$ are supersymmetric gauge theories. If, instead, $l\gg l_s$
then we cannot ignore interactions due to brane fluctuations but we can ignore
inter-brane interactions. In this limit the dynamics of each brane is governed
(at sufficiently low energy) by a Dirac-Born-Infeld (DBI) action. At intermediate
length scales we have interactions of both types. This regime is the one that
we are going to study in these lectures, although not for D-branes. We want to
include the inter-brane interactions that can lead to interacting field
theories, but we want to go beyond the field theory approximation by including
the interactions due to large brane fluctuations. We shall call this `brane
theory'.

One might expect some general features of field theory to remain valid in brane
theory, and others not. Of importance to these lectures is the fact that many
supersymmetric field theories admit supersymmetric soliton solutions
saturating a Bogomolnyi-type bound. As these bounds are usually a consequence
of the supersymmetry algebra one would expect them to hold beyond the field
theory approximation, but it is not immediately clear how, or whether, the bound
continues to be saturated because the brane theory equations are different. In
the D-brane case, for example, the equations are of non-abelian  DBI type. They
reduce to standard gauge theory equations in the field theory limit, with their
standard gauge theory soliton solutions, but to determine whether these
solitons continue solve the brane theory equations and, if so, whether they
continue to saturate the energy bound implied by supersymmetry requires
a precise knowledge of the non-abelian DBI equations. Here we confront a
general difficulty: to get interactions of field theory type we need more than
one brane, but the inter-brane interactions are known precisely only in the
field theory limit. 

There is one way to avoid this dilemma. We can fix our attention on one brane 
and replace the others with which it interacts by the supergravity background
that they induce at `our' brane. This way we take the inter-brane interactions
into account in an approximation that is exact in the limit of large
brane charges for the `other' branes, which are effectively
macroscopic and can be replaced by the supergravity background
they induce. What we now have is a single brane in a brane
background. As applied to
D-branes, this method can be used to recover many of the finite-energy soliton
solutions of D=4 SYM theory but as soliton solutions on a {\sl single} brane,
with an {\sl abelian} gauge group \cite{Marija}. In this application, the `other'
branes are parallel D3-branes. Another application of this idea, and one to be
reviewed in lecture 2, involves M2-branes in a two-centre M-monopole
(i.e. M-theory Kaluza-Klein monopole) background, which
allows a brane realization \cite{BT} 
of the sigma model `lump' soliton of hyper-K\"ahler
sigma-models. Of course, other backgrounds can yield brany
generalizations of sigma-models  with other target spaces, and other branes can
yield models in other dimensions. A general feature is that the relativitistic
aspects of the field theory are extended to a bulk-space relativity. This is
explained in the first lecture, where it is presented as a consequence of
implementing the principle of `field-space democracy' \cite{albert}. 

It is also a general feature that sigma-model solitons survive as solutions of
the brane equations and continue to minimize the energy in their charge
sector. Moreover, these solitons now acquire a new geometrical interpretation, 
as minimal surfaces in the simple cases discussed in these lectures. At this
point we can see that brane theory goes beyond field theory because there are
many types of minimal surface and not all of them have an interpretation as
field theory solitons. Derrick's theorem states that static minimal energy
solutions of (conventional) scalar field theories in $p$ space dimension 
cannot exist for $p>2$, but Derrick's theorem no longer applies once we 
have made the transition from field theory to brane theory. 

When discussing solitons that minimise the energy it is natural to start from
the Hamiltonian rather than the Lagrangian, and this will be the strategy
adopted here, following \cite{GGT}. Only the standard Dirac-type p-brane action,
and its Hamiltonian, will be needed in these lectures, apart from the coupling
to a background (p+1)-form gauge potential $A$, which will play a minor role
because only backgrounds with vanishing field strength $F=dA$ will arise. Neither
will we need terms involving worldvolume gauge fields because no attempt will be
made here to review the brane theory status of {\sl gauge theory} solitons.
A further restriction will be to {\sl static} solitons. A general framework for
those cases that remain is provided by the theory of calibrations \cite{HL}. As
we shall see in the third lecture, sigma model solitons on the M2-brane fit into
this framework as  examples of K\"ahler calibrations \cite{GPT}, but for other
branes there are solitons with no field theory analogue that arise from more
complicated types of calibration \cite{BBS,oog}. The simplest example, albeit 
one with infinite energy, is the Special Lagrangian 3-surface in $\bE^6$. Its
realization by intersecting M5-branes \cite{GibPap,GLW,QMW} will be reviewed in
the fourth, and last, lecture. Hopefully, the detailed treatment of these
few cases of `brane theory solitons' will compensate for the restricted focus.

One other restriction is implicit in the above discussion. We have
taken the term `field theory' to exclude gravity (and hence supergravity). It
has long been appreciated that gravity is a rather special kind of `field
theory', and branes have provided us with a new reason for believing this.
Gravitons (and superpartners) propagate in the `bulk' while `matter' propagates 
on branes. There are reasons for this that go beyond the simple statement that
it is `difficult' to confine gravity to a brane but these reasons need to be
reassessed in light of the demonstration \cite{RS} that it is, nevertheless,
possible. The epilogue that concludes these lectures deals with some of these
issues. 

\section{Lecture 1: Field theory vs Brane Theory}

The $n$ scalar fields of a non-linear sigma-model define a map from a
(p+1)-dimensional Minkowski spacetime $W$ with metric $\eta$ (diagonal with
entries $(-1,1,\dots,1)$ in cartesian coordinates) to an
$n$-dimensional Riemannian target space $M$ with metric $G$. Let $\xi^\mu$,
($\mu=0,1,\dots, p$) be {\sl cartesian} coordinates for $W$, and let $X^i$ 
($i=1,\dots,n$) be  coordinates for $M$, so that $X^i(\xi)$ are the scalar
fields. The Lagrangian density of the massless sigma model is then
\be\label{lagden}
{\cal L}_\sigma = {1\over2}\eta^{\mu\nu}\partial_\mu X^i 
\partial_\nu X^j G_{ij}(X) \, .
\ee
The corresponding Hamiltonian density is
\be\label{hamden}
{\cal H}_\sigma = {1\over2}G^{ij}P_iP_j + {1\over2}\bfn
X^i\cdot \bfn X^j G_{ij}\, ,
\ee
where the variables $P_i(\xi)$ are the momenta canonically conjugate to the
fields $X^i(\xi)$. Although this field theory is `relativistic' as a field
theory on $W$, there is another sense in which it is {\sl not} relativistic.
Consider the special case of $p=0$, and set $P_i = p_i/\sqrt{\mu}$; then
${\cal H}_\sigma = H_{NR}$, where
\be 
H_{NR} = {1\over2\mu}G^{ij}p_ip_j \, .
\ee
This is the Hamiltonian of a {\sl non-relativistic} 
particle of mass $\mu$. This
may be contrasted with the Hamiltonian 
\be\label{relpart}
H = \sqrt{G^{ij}p_ip_j + \mu^2}
\ee
for a {\sl relativistic} particle. Is there an analogous `{\sl relativistic}'
Hamiltonian density ${\cal H}$ for a scalar field theory? 

There is, and one way to find it is by implementation 
of `field-space democracy'
\cite{albert} (a principle invoked for similar reasons in
\cite{azco}). 
We begin with a Lorentzian spacetime ${\cal M}$ of
dimension $D=p+n+1$ with coordinates $(\xi^\mu,X^i)$. The sigma model
fields $X^i(\xi)$ now define a (p+1)-dimensional surface $W$ in ${\cal M}$. 
This (p+1)-surface can also be specified parametrically by giving $D$
coordinates $X^m$ ($m=0,1,\dots, p+n$), as functions of $(p+1)$ parameters
$\xi^\mu$, such that $X^\mu= \xi^\mu$ for some particular parametrization. 
The equations for the (p+1)-surface $W$ now take the form $X^m=X^m(\xi)$. We
will assume that the induced metric on $W$ is Lorentzian, so that $W$ is
a (p+1)-dimensional worldvolume swept out in time by a p-dimensional
`worldspace' $w$. The evolution of $w$ is governed by an action $S[X]$ that
must be reparametrization invariant in order to allow, at least locally, the
`physical' gauge choice 
\be\label{physgauge}
X^\mu(\xi)=\xi^\mu\, .
\ee
Let ${\cal G}$ be the Lorentzian metric on ${\cal M}$ and let $g$ be the metric
it induces on $W$. The reparametrization invariant action with a Lagrangian
density of lowest dimension is
\be\label{pact}
S = -T\int d^{p+1}\xi \, \sqrt{-\det g}\, ,
\ee
which is the Dirac-Nambu-Goto action for a p-brane of tension $T$. 

To make contact with the sigma model we set $T=1$ and take the metric 
${\cal G}$ on ${\cal M}$ to be of the form
\be\label{block}
{\cal G}_{mn} = \pmatrix{\eta_{\mu\nu} & 0 \cr 0 & G_{ij}}\, .
\ee
The physical gauge metric that this induces on $W$ is
\be\label{inducedm}
g_{\mu\nu} = \eta_{\mu\nu} + \partial_\mu X^i \partial_\nu X^j G_{ij}\, .
\ee
Choosing local {\sl cartesian} coordinates for $W$, we then have 
\be
T^{-1}{\cal L} = -\det g = 1 + 
{1\over2} \eta^{\mu\nu}\partial_\mu X^i \partial_\nu X^j
G_{ij} + {\cal O}\left((\partial X)^4\right)\, .
\ee
Apart from the constant term, and the higher-derivative corrections, this is
the sigma-model Lagrangian density. The low-energy `non-relativistic' dynamics 
of the brane is therefore governed by the sigma-model. To complete the picture 
we now need to determine the p-brane Hamiltonian and show that it provides the
required generalization of the relativistic particle hamiltonian. 

We could find the physical-gauge Hamiltonian by performing a Legendre
transformation on the gauge-fixed Lagrangian. Instead, we will first proceed to
the Hamiltonian form of the gauge-invariant action. This has the advantage of
maintaining manifest invariance under any isometries of ${\cal G}$, until
a gauge choice is made. We will need to make a worldvolume space/time split, so
we write
\be
\xi^\mu=(t,\sigma^a)\, ,
\ee
where $\sigma^a$ ($a=1,\dots,p$) are coordinates for the $p$-dimensional
worldspace $w$. Let us write the induced metric on $W$ as
\be
g_{\mu\nu} =\pmatrix{g_{tt}&g_{ta}\cr g_{tb}&m_{ab}}\, ,
\ee
so that $m$ is the metric induced on $w$, with components
\be
m_{ab} =\partial_a X^m\partial_b X^n {\cal G}_{mn}\, .
\ee
One can now use standard methods to obtain the Hamiltonian form of the action in
which the independent variables are the scalar fields $X^m(\xi)$ and their
canonically conjugate momenta $P_m(\xi)$. The result is \cite{BSTanii}
\be\label{ham}
S= \int dt \int d^p\sigma \, \left[ \dot X^m P_m - 
s^i {\cal H}_i - \ell {\cal H}_t\right]\, ,
\ee
where $s^a$ and $\ell$ are Lagrange multipliers for the constraints ${\cal
H}_a=0$ and ${\cal H}_t=0$, with
\be
{\cal H}_a = \partial_a X^m P_m \, ,\qquad
{\cal H}_t = {1\over2} \left({\cal G}^{mn}P_mP_n + T^2\det m\right)\, .
\ee
This form of the action is to be expected from the
general covariance of the initial action (\ref{pact}). The Lagrange multipliers
are analogous to the `shift' and  `lapse' functions of General Relativity.
The difference is that the geometry here is {\sl extrinsic} whereas that of
General  Relativity is {\sl intrinsic}. 

It is simple to verify that the result given above is correct. Elimination of
$P_m$ in (\ref{ham}) by its Euler-Lagrange equation yields
\be
S= \int dt \int d^p\sigma \,\left[{1\over 2\ell}\left(g_{tt} - 
2s^ag_{ta} + s^a s^b m_{ab}\right) -{1\over2} T^2\ell \det m\,\right] .
\ee
We now eliminate $s^a$ by its Euler-Lagrange equation, set
\be
\ell = v/\det m\, ,
\ee
and use the identity 
\be\label{detiden}
\det g \equiv \det m\left( g_{tt}- m^{ab}g_{ta}g_{tb}\right)\, ,
\ee 
to get
\be
S= \int dt \int d^p\sigma\,\left[ {1\over 2v}\det g -{1\over2} T^2 v\,\right] .
\ee
This is a well-known alternative form of the p-brane action.
Provided $T\ne0$, which we assume here, we can eliminate $v$ by its algebraic
Euler-Lagrange equation to recover (\ref{pact}). 
 
To find the physical gauge Hamiltonian we have only to substitute the 
`physical' gauge choice (\ref{physgauge}) into the constraints and then solve
them for the momenta $P_\mu$. However, it is instructive to proceed 
sequentially, first fixing only the time parameterization by the gauge choice
$X^0(\xi)=t$. If we rename $P_0$ as $-{\cal H}$, and define
\be
X^I=(X^a,X^i) \, ,\qquad P_I=(P_a, P_i)\, ,
\ee
then we now have
\be
X^m=(t,X^I) \, ,\qquad P_m=(-{\cal H},P_I)\, .
\ee
It will also prove convenient to write the spacetime metric ${\cal G}$ as
\be
{\cal G}_{mn} = \pmatrix{G_{00}& G_{0I}\cr G_{0J} & M_{IJ}}\, .
\ee
Note that since $\partial_a X^0 =0$ the metric $m$ is now
\be\label{minduced}
m_{ab} = \partial_a X^I \partial_a X^J M_{IJ}\, .
\ee

The Hamiltonian constraint ${\cal H}_t=0$ can now be solved to yield
\be\label{hamsol}
{\cal H} = {\cal N}^I P_I \pm 
{\cal N} \sqrt{ M^{IJ} P_IP_J + T^2 \det m}\, ,
\ee
where
\be
{\cal N}^I= -{{\cal G}^{0I}\over {\cal G}^{00}}\, , 
\qquad {\cal N} = {1\over \sqrt{-{\cal G}^{00}}}\, ,
\ee
and $M^{IJ}$ is the inverse of the space metric $M_{IJ}$.
The action (\ref{ham}) now becomes
\be
S= \int dt \int d^p\sigma \, \left[ \dot X^IP_I - {\cal H}(X,P) - s^a {\cal
H}_a\right]\, ,
\ee
which is that of a $(p+1)$-dimensional field theory with Hamiltonian density
${\cal H}(X,P)$. The constraint imposed by $s^a$ is {\sl linear} 
in momenta and 
can therefore be viewed as the generator of a gauge invariance.  

To fix this gauge invariance we set $X^a(\xi)= \sigma^a$. 
The constraint ${\cal H}_a=0$ can then be solved for $P_a$,
\be
P_a =-\partial_a X^iP_i\, ,
\ee
and the action then takes the canonical form
\be
S= \int dt \int d^p\sigma \, \left[ \dot X^iP_i - {\cal H}\right]\, ,
\ee
where the Hamiltonian density is now a function only of the
physical phase-space variables $(X^i,P_i)$. 

For a metric on ${\cal M}$ of the
form (\ref{block}) we  have ${\cal N}^I=0$, ${\cal N}=1$ and
\be\label{Mform}
M_{IJ} = \pmatrix{\delta_{ab}&0\cr 0& G_{ij}}\, .
\ee
The physical-gauge metric on $w$ is therefore
\be\label{mform}
m_{ab} = \delta_{ab} + \partial_a X^i \partial_b X^j G_{ij}\, .
\ee
The Hamiltonian density for this case is 
\be\label{hamdensitya}
{\cal H} = \sqrt{\left(G^{ij} + \bfn X^i \cdot \bfn X^j\right) P_iP_j
+ T^2 \det\left( \bI + \bfn X^i \bfn X^j G_{ij}\right)}\, .
\ee
For $p=0$ the $\bfn X$ terms are absent and $T=\mu$, a mass parameter. The
Hamiltonian density then reduces to the Hamiltonian (\ref{relpart}) for a
relativistic particle; we have thus found the sought $p>0$ generalization of
this Hamiltonian. If we now set $T=1$ and write (\ref{hamdensitya}) as a double
expansion in  powers of $P$ and $\bfn X$, we find that
\be\label{cale}
{\cal H} = 1 + {1\over2}\left[ G^{ij}P_iP_j + \bfn X^i \cdot
\bfn X^j G_{ij}\right] + \dots
\ee
The leading term is the p-surface tension energy of the brane. 
The next term is just the sigma model hamiltonian. The remaining terms,
indicated by the dots, are `relativistic' corrections; these can be ignored
if (i) all speeds are much less than light, {\sl and} (ii) all fields are slowly
varying. To this we should add that the validity of the Dirac-Nambu-Goto action
from which we began requires all accelerations to be small. 

Although the non-zero vacuum energy is expected, it is natural to define the
energy on any given worldspace as 
\be
{\cal E} = {\cal H} -T\, ,
\ee
because this vanishes in the vacuum. Although no mention has been
made of supersymmetry, it is nevertheless the case that the analogous analysis
for a super-p-brane action yields a supersymmetric worldvolume theory for which
the energy density must vanish in the vacuum. In fact, it is ${\cal E}$,
rather than ${\cal H}$, that plays the role of the Hamiltonian density in the
worldvolume supersymmetry current algebra \cite{HP}, and this is what allows
the brane vacuum to preserve half of the supersymmetry of the spacetime vacuum.
This can also be understood, from the spacetime perspective, as due to a 
p-form charge in the spacetime supersymmetry algebra \cite{AGIT}. For these
reasons, we will usually focus on the worldspace energy density ${\cal E}$
given in the spacetimes of interest here, and for $T=1$, by the formula
\be\label{hamdensity}
({\cal E} +1)^2 = \left(G^{ij} + \bfn X^i \cdot \bfn X^j\right) P_iP_j
+ T^2 \det\left( \bI + \bfn X^i \bfn X^j G_{ij}\right)\, .
\ee

\section{Lecture 2: Sigma-model solitons on branes}

Let us begin with the (2+1) dimensional sigma-model Hamiltonian
\be\label{hamsig}
H_\sigma = {1\over2}\int d^2\sigma\, \{|P|^2 + |\bfn X|^2\}\, ,
\ee
where the norm $|.|$ is defined by contraction with the target space metric
and, where applicable, with the Euclidean 2-space metric. We use here standard
vector calculus notation for differential operators on $\bE^2$. For example, in
cartesian coordinates we have 
\be
\bfn = (\partial_1,\partial_2), \qquad
\star\bfn = (\partial_2,-\partial_1)\, .
\ee
We will not consider models with fermions, such as supersymmetric
models. However, all the models we will consider are supersymmetrizable, so
that supersymmetry will be implicit in much of the discussion and it
will pay to keep in mind some of its implications.
The simplest, N=1, (2+1)-dimensional 
supersymmetric sigma model has one real $Sl(2;\bR)$ spinor
charge. If the target space has a metric of reduced holonomy then there may
be additional supersymmetries. Specifically, if the target space is
K\"ahler then there will be two spinor charges \cite{zumino} ($N=2$
supersymmetry) and if it is hyper-K\"ahler there will be four
spinor charges \cite{AGF} (N=4  supersymmetry). A summary of what these 
K\"ahler and hyper-K\"ahler conditions mean now follows.

If the target space $M$ is almost-complex then it will admit an almost complex
structure, which is a (1,1) tensor $I$ such that $I^2=-\bI$, 
where $\bI$ is the identity matrix. 
Given an almost-complex structure $I$ we may define the associated Nijenhuis
tensor
\be
N_{ij}{}^k(I) = 4\left( \partial_\ell I_{[i}{}^k I_{j]}{}^\ell + \partial_{[i}
I_{j]}{}^\ell I_\ell{}^k \right)\, .
\ee
If $N(I)$ vanishes then $I$ is a complex structure and $M$ is a complex 
manifold. A metric $G$ on $M$ satisfying
\be
I_{(i}{}^j G_{j)k} =0
\ee
is Hermitian with respect to $I$. For a Hermitian metric the
tensor $I_{ij}$ is antisymmetric and hence defines a 2-form 
\be
\Omega = {1\over2}I_{ij}\, dX^i\wedge dX^j\, .
\ee
The metric is K\"ahler if this 2-form is closed, $d\Omega=0$, and
$\Omega$ is then called the K\"ahler 2-form (associated to the complex
structure $I$). For a complex manifold, with vanishing Nijenhuis tensor, this
condition is equivalent to the apparently weaker condition that $I$ be
covariantly constant (with respect to the usual affine metric connection). A
hyper-K\"ahler manifold is one with a metric that is K\"ahler with respect to
three independent complex structures
$I,J,K,$ obeying the algebra of the quaternions ($IJ=K$ and cyclic).

We begin our study of solitons by seeking minimal energy configurations 
of a K\"ahler sigma model. The Hamiltonian (\ref{hamsig}) can be rewritten as
\be\label{bog}
H_\sigma = {1\over 4}\int d^2\sigma\, \big\{2|P|^2 + 
|\bfn X \mp \star\bfn X I|^2\big\}\,  \mp\, L .
\ee
where $L$ is the topological `lump' charge
\be
L= \int_w \Omega\, .
\ee
The integrand is the the K\"ahler 2-form $\Omega$, which is integrated over the
2-surface $w$ into which the Euclidean 2-space is mapped by the sigma models
map.  To check the equivalence of (\ref{bog})
to the original form (\ref{hamsig}) it suffices to note that $L$
cancels against the cross term from 
\be
G_{ij}\left(\bfn X^i \mp \star\bfn X^k
I_k{}^i\right) \left(\bfn X^j \mp \star\bfn X^l I_l{}^j\right)\, ,
\ee
while the identity $I_{(k}{}^i I_{\ell)}{}^j G_{ij} = G_{k\ell}$
ensures equality of the remaining two terms. 

Since $L$ is a topological invariant, the variation of the fields for fixed
boundary conditions will not change its value, and since the other
terms in $H$ are non-negative we deduce the bound \cite{per}
\be
H_\sigma \ge |L|\, ,
\ee
which is saturated by {\sl static} solutions of the first order equations
\be\label{CR}
\bfn X^i =\pm \star\bfn X^k I_k{}^i\, .
\ee
Locally we may choose complex coordinates $Z^\alpha$ on a chart of $M$ for 
which
$I$ is diagonal with eigenvalues $\pm i$. We may also view $\bE^2$ as the complex
plane with complex coordinate $\zeta=\sigma_1 \pm \sigma_2$. The equations
(\ref{CR}) then reduce to
\be
\bar\partial Z^\alpha =0\, ,
\ee
where $\bar\partial\equiv \partial/\partial \bar\zeta$. That is, the functions
$Z^\alpha(\zeta)$ are holomorphic functions. Globally this means that the
solutions of (\ref{CR}) are holomorphic curves on $M$. 

As a simple example, suppose that $M=\bC$, with complex coordinate $Z$ and flat
metric $dZd\bar Z$. The complex sigma-model field is $Z(t,\zeta,\bar
\zeta)$, but as there are no interactions we can hardly expect to find
solitons. All the same, it will prove instructive to consider how one might go
about looking for them. As we have seen, static solutions of minimum energy
correspond to holomorphic functions $Z(\zeta)$. For a localized
energy density we require that $|Z|\rightarrow 0$ as 
$|\zeta|\rightarrow \infty$.
This means that any non-zero $Z(\zeta)$ must have singularities, and the 
simplest choice is a point singularity at the origin. For this choice we have
\be
Z(\zeta)= c/\zeta
\ee
for complex constant $c$, with $|c|$ determining the objects's `size'. It would
be misleading to call this object a `soliton' because its energy is infinite.
To see this we note that the Kahler 2-form on $\bC$ is $\Omega = idZ
\wedge d\bar Z$ so its pullback to the complex $\zeta$-plane, when $Z$ is
holomorphic has magnitude $|Z'|^2$. The soliton energy is therefore
\bea
E &=& \int d^2\sigma |Z'|^2 = |c|^2 \int d^2\sigma |\zeta|^{-4}\\
&=& -\pi |c|^2 [r^{-2}]_0^\infty = [\pi R^2]_0^\infty\, ,
\eea
where $r$ is distance from the origin in the $\zeta$-plane, and $R$ is distance
from the origin in the $Z$-plane. The energy is infinite because it equals the
infinite area of the target 2-space. In general, a finite energy soliton
saturating the energy bound is possible only if the target space has a compact
holomorphic 2-cycle. A holomorphic map $Z^\alpha(\zeta)$ then yields finite
energy if it maps the $\zeta$-plane to this 2-cycle, and the  energy will be the
area of the 2-cycle. Obviously, a flat target space, which yields a free field
theory, does not have such 2-cycles. 

A flat target space has trivial holonomy. In some respects, the simplest 
non-flat sigma models are those for which the holonomy group is the smallest
non-trivial subgroup of $SO(n)$. If one also requires a Ricci flat metric (this
being motivated by its ultimate interpretation as part of a background
supergravity solution) then the simplest case is $n=4$ with holonomy
$SU(2)\subset SO(4)$. Such 4-manifolds are hyper-K\"ahler. In this case there
is a  triplet ${\bf I}$ of complex structures. For any unit 3-vector
${\bf n}$ the tensor $I= {\bf n}\cdot {\bf I}$ is also a complex structure,
which we can identify as the one of the above discussion. Similarly,
$\Omega= {\bf n}\cdot \bfO$, where $\bfO$ is the triplet of 
K\"ahler 2-forms. An important class of hyper-K\"ahler 4-manifolds are those
admitting a tri-holomorphic Killing vector field; that is, a Killing vector
$k$ field for which ${\cal L}_k \bfO$ vanishes.  All such manifolds are 
circle bundles over $\bE^3$ \cite{gibrub}. We can choose coordinates such that
\be
k=\partial/\partial\varphi\, ,
\ee
where $\varphi$ parametrizes the circle. The metric then takes the form 
\be\label{gibhawk}
ds^2_4 = V^{-1} (d\varphi-{\bf A}\cdot d{\bf X})^2 + Vd{\bf X}\cdot d{\bf X}
\ee
where $\bfn \times {\bf A}=\bfn V$. This implies that $V({\bf X})$ is harmonic
on $\bE^3$, except at isolated poles. The metric (\ref{gibhawk}) is complete
provided that (i) the residues of $V$ at its poles are equal and positive, and
(ii) $\varphi$ is an angular variable with period $4\pi$ times this common
residue. Under these circumstances the poles of $V$ are coordinate singularities
of the metric, called its `centres'. If we also take $V\rightarrow 1$ as $|{\bf
X}|\rightarrow \infty$ then the metric is asymptotically flat. 
A simple example is the 2-centre metric with $\varphi \sim \varphi +
2\pi$ and
\be
\label{twocentre}
V=1+ {1\over2}\left[{1\over |{\bf X}+ {\bf a}|} + 
{1\over |{\bf X}- {\bf a}|}\right]\, .
\ee

In terms of the frame 1-forms
\be
e^\varphi = V^{-{1\over2}}\left(d\varphi - {\bf A}\cdot d{\bf X}\right)\, ,
\qquad {\bf e} = V^{1\over2}d{\bf X}\, ,
\ee
the triplet of K\"ahler 2-forms is 
\be
\bfO = e^\phi{\bf e} - {1\over2} {\bf e}\times {\bf e}\, ,
\ee
where the wedge product of forms is implicit here. 
In the two-centre case there is a
preferred direction ${\bf n} = {\bf a}/|{\bf a}|$ and hence a preferred complex
structure $\Omega={\bf n}\cdot \bfO$. The 2-centre metric is the
simplest multi-centre metric, all of which admit finite energy lump solutions
corresponding to holomorphic maps from $\bC$ to homology 2-cycles. 
In the 2-centre case there is just one such 2-cycle. This
is the 2-sphere with poles at the centres, where $k$ vanishes, and orbits of
$k$ as its lines of latitude. The lump solution can be found from the ansatz
${\bf X}= X{\bf n}$, which leaves leaves $\varphi(\bfs)$ and $X(\bfs)$ as the 
two `active' coordinates. When restricted to this subspace, the K\"ahler
2-form is $\Omega = d\varphi\wedge dX$, and hence $|L| = 4\pi |{\bf a}|$. 

We now wish to generalize these considerations from field theory to brane
theory. Our starting point will be the $p=2$ case of the formula
(\ref{hamdensity}) for the  physical-gauge p-brane energy density ${\cal E}$.
We expand the $2\times 2$ determinant to obtain
\bea\label{hamdenbrane}
({\cal E}+1)^2 &=& 1 + |\bfn X|^2 + (G^{ij} + \bfn X^i\cdot \bfn X^j)P_iP_j\nn
&& +\ 2 X^{ij}X^{kl}G_{ik}G_{jl}\, ,
\eea
where we have set
\be
X^{ij} \equiv {1\over2}\bfn X^i \times\bfn X^j\, .
\ee
Previously we were able to express the energy as a sum of a topological
charge and a manifestly non-negative integral. This is the trick introduced by
Bogomol'nyi for deriving energy bounds in field theory \cite{bogomol}. Its
generalization to brane theory involves writing $({\cal E}+1)^2$ as a sum of
squares \cite{HL,GGT,BT}. To simplify we will put the 
momentum to zero. For the case in hand 
we can then rewrite (\ref{hamdenbrane}) as 
\bea
({\cal E}+1)^2 &=& \left(1 \mp X^{ij}I_{ij}\right)^2 + 
{1\over2}|\bfn
X \mp \star\bfn X I|^2 \nn  && + \left(X^{ij}J_{ij}\right)^2 
+ \left(X^{ij}K_{ij}\right)^2 \, .
\eea
To verify this one needs the identity
\be
\delta_i{}^{(j} \delta_l{}^{k)} + I_i{}^{(j}I_l{}^{k)} + J_i{}^{(j}J_l{}^{k)} +
K_i{}^{(j}K_l{}^{k)} \equiv G^{jk}G_{il}\, .
\ee
It now follows (for one choice of sign) that
\be
{\cal E} \ge  |X^{ij}I_{ij}|\, .
\ee
This bound is saturated by static
solutions of the same first-order equations (\ref{CR}) as we found before
because, for example, these imply that
\bea
X^{ij}J_{ij} &=& \mp {1\over 2} \bfn X^i \cdot \bfn X^j (IJ)_{ij}\nn
&=& \mp {1\over 2} \bfn X^i \cdot \bfn X^j K_{ij} \equiv 0\, ,
\eea
where we have used $IJ=K$ in the last line. Of course, the choice of complex
structure $I$ is arbitrary; we could take $I={\bf n}\cdot {\bf I}$. Let
\be
{\bf L} = \int_w\!\bfO
\ee
and let $\bar{\bf n}$ be the direction that minimises ${\bf n}\cdot {\bf L}$.
Then we deduce the bound
\be\label{branebound}
{\cal E} \ge  |X^{ij}\Omega_{ij}|\, ,
\ee
where $\Omega = \bar{\bf n}\cdot \bfO$. Integration of (\ref{branebound})
yields the bound 
\be\label{branebound2}
E \equiv \int d^2\sigma\, {\cal E} \ge |L|
\ee
on the total energy. Given that the bound (\ref{branebound}) is
saturated, the bound (\ref{branebound2}) will also be saturated {\sl 
provided that the integrand of $L$ does not change sign}. 
Recalling that $\Omega_{ij}\equiv I_{ij}$, we see that this condition
is satisfied because (\ref{CR}) implies that
\be
X^{ij}\Omega_{ij} = \pm |\bfn X|^2\, .
\ee
Thus, the bound (\ref{branebound2}) is saturated by static solutions of 
(\ref{CR}). We see that the additional non-quadratic terms in the
membrane Hamiltonian make no difference to the final result. 

Let us now reconsider the case in which $M=\bC$. This corresponds to a membrane
in a 5-dimensional Minkowski spacetime, which we can view as the product of a
real time-line with $\bC^2$; the $\bC^2$ coordinates are $(Z,\zeta)$. Minimal
energy membranes are static holomorphic curves in $\bC^2$, which are specified
by an equation of the form $f(Z,\zeta)=0$ for some holomorphic function of $Z$
and $\zeta$. This equation has a solution of the form $Z=Z(\zeta)$ in which
$\zeta$ parametrizes the membrane worldspace $w$ and $Z(\zeta)$ can now be
interpreted both as a worldvolume field and as the displacement of the membrane
in the $Z$-plane at the coordinate $\zeta$. If we want the field $Z(\zeta)$ to 
be single-valued on the $\zeta$-plane, a condition that is normally required of
a sigma model, we must choose $f$ to be linear in $Z$. In contrast, we might
expect any given value of $Z$ to occur for several values of $\zeta$; for
instance, if we have $k$ identical widely-separated solitons we expect each
value of $Z$ to occur at least $k$ times. This will happen if $f$ is a
$k$'th order polynomial in $\zeta$, but this suggests that the one soliton
sector is described by a function $f$ that is {\sl linear} 
in $\zeta$. The simplest
soliton solution should therefore be found by choosing $f= \zeta Z -c$.
Provided that $c\ne 0$, this yields the solution $Z(\zeta) = c/\zeta$ discussed
above. In that discussion the limit $c\rightarrow 0$, which shrinks the
`soliton' to a point, would simply yield the sigma-model vacuum $Z\equiv 0$.  
In the membrane context, however, this limit yields the equation
\be\label{intersect}
\zeta Z=0\, ,
\ee
which has two solutions: $Z=0$ or $\zeta=0$. The second solution makes no sense
in the sigma model context but it does in the membrane context. Since
this equation is symmetric under the interchange of $Z$ and $\zeta$ we could
equally well interpret $Z$ as a worldspace coordinate and $\zeta(Z)$ as its
displacement in the $\zeta$-plane. Thus, the equation (\ref{intersect})
describes two membranes intersecting at the point $Z=\zeta=0$. Recalling that
this is a limit of the equation
\be
\zeta Z =c
\ee
we see that the `soliton' solution $Z=c/\zeta$ describes the desingularized
intersection of two membranes \cite{CMG}. Either membrane can be viewed as an
infinite-energy `soliton' on the worldspace of the other one, the energy being
infinite because the `soliton' membrane has constant surface tension and infinite
area. Of course, it is also possible to view the desingularized intersection as
a {\sl single} membrane in $\bE^4$ with two asymptotic planes. This single
membrane will have minimal energy if it is a minimal surface in $\bE^4$. The
study of sigma model solitons is therefore closely related to the study of
minimal surfaces. We shall return to this theme in the next two lectures. 

We now turn to the membrane version of the finite energy soliton of the
hyper-K\"ahler sigma model. We start from the D=11 supermembrane 
in a D=11 supergravity M-monopole background. The supermembrane is a super
version of the membrane already considered, and can be consistently formulated
in any background that solves the D=11 supergravity field equations
\cite{BSTown}. The M-monopole is a solution of D=11 supergravity \cite{han}
for which the only non-vanishing field is the 11-metric, which takes the form 
\be
\label{m11}
ds_{11}^2= ds^2(\bE^{(1,6)}) + G_{ij}(X)dX^idX^j\, ,
\ee
where $G$ is a hyper-K\"ahler 4-metric of the type considered above. 
We now place a probe membrane in this background and choose its vacuum to be a
Minkowski 3-space in $\bE^{(1,6)}$. Restricting attention to deformations of the
supermembrane described by the worldvolume fields $X^i$, we find the induced
worldvolume 3-metric to be exactly as in (\ref{inducedm}). As we have seen, this
leads, in the field theory limit, to a sigma-model with target space metric
$G$. In the case of the supermembrane this becomes an N=4 supersymmetric sigma
model. As we have seen this model admits finite energy (and 1/2 supersymmetric)
lump solutions corresponding to particular holomorphic curves. 
We have also seen
that the {\sl same} configurations minimise the brane theory energy, with the
membrane worldspace as the holomorphic curve. For the choice of a 2-centre
hyper-K\"ahler metric with $V$ given by (\ref{twocentre})  we have the finite
area homology 2-sphere previously described and the lump is a membrane wrapped
on it. This appears as a soliton on a probe brane that intersects the lump
brane. Non-singular intersections, which can be viewed as {\sl single}
membranes asymptotic to the vacuum membrane,  are obtained as solutions of
(\ref{CR}). This entire set up can be summarized by the array
$$
\begin{array}{lcccccccccccc}
MK: &   &   & \times  & - & - & - &   &   &   &   &   &   \nn
M2: & 1 & 2 &   &   &   & - &   &   &   &   &   &   \nn
M5: &   &   & 3 & 4 &   &   &   &   &   &   &   &  
\end{array}
$$ 
where `MK' indicates the (multi-centre) M-monopole background solution of D=11
supergravity; the cross represents the compact direction of this
background. The second row is the probe supermembrane, or M2-brane, and the 
third row the soliton M2-brane; of course, in the case of a non-singular
intersection, there is really only one M2-brane. This array is associated with
the constraints
\be\label{memconstraint}
\Gamma_{3456}\epsilon =\epsilon\, , \qquad \Gamma_{012}\epsilon =\epsilon\, ,
\qquad \Gamma_{034}\epsilon=\epsilon\, ,
\ee
where $\epsilon$ is a 32-component real D=11 spinor. I refer to my previous
Carg\`ese lectures \cite{carg} for an explanation of these constraints, which
will be needed in the following lecture.  The fact that their solution space is
is 4-dimensional implies that the configuration as a whole preserves
1/8 of the supersymmetry of the M-theory vacuum. As the hyper-K\"ahler sigma
model vacuum preserves eight supersymmetries the lump soliton on the
supermembrane preserves 1/2 of the supersymmetry
of the brane theory vacuum. 

To conclude this lecture, we will consider the IIA interpretation of the above 
sigma 
model lump.  Because of the holomorphicity of $k$, reduction on its orbits
preserves all supersymmetries of the original configuration. I will not prove
this here, but it can be verified directly from the resulting IIA
configuration, which (after a permutation of the columns) is represented by
the array
$$
\begin{array}{lccccccccccc}
D6: & 1 & 2 & 3 & 4 & 5 & 6 &   &   &   &   &   \nn
D2: & 1 & 2 &   &   &   &   &   &   &   &   &   \nn
F1: &   &   &   &   &   &   & 7 &   &   &   &    
\end{array}
$$ 
We now have two parallel D6-branes, represented by the first row. The probe
M2-brane has become a D2-brane parallel to the D6-branes and the `soliton'
M2-brane a IIA string stretched between the D6-branes. An intersection
of the string with the D2-brane corresponds to a singular intersection of the
two M2-branes. The deformation of the M2-branes to a non-singular lump on a
single M2-brane now has a IIA interpretation as the splitting of the IIA string
intersection with the D2-brane into two endpoints, yielding two separate IIA
strings stretched between the D2-brane and each of the D6-branes.

\section{Lecture 3: Solitons and K\"ahler Calibrations}

We have been considering p-branes in D-dimensional spacetimes ${\cal M}=\bR
\times {\cal S}$ with metric
\be
ds^2 = -(dx^0)^2 + M_{IJ}dX^IdX^j\, ,
\ee
so that $M$ is the Riemannian metric on the (D-1)-dimensional space ${\cal S}$.
We shall assume that a time parametrization has been chosen so that
$X^0(\xi)=t$. A static p-brane is then an immersed p-surface $w$ in ${\cal S}$
specified by functions $X^I(\xi)$. The metric induced on $w$ is 
the metric $m$ of (\ref{minduced}). 
Let $\Gamma_I$ be the spatial Dirac matrices that anticommute with
$\gamma_0$ and satisfy
\be
\{\Gamma_I,\Gamma_J\} =2M_{IJ}\, ,
\ee
and let $\Gamma_{IJ\dots}$ be antisymmetrized products of Dirac matrices (with
`strength one', so that $\Gamma_{12\dots}=\Gamma_1\Gamma_2\cdots$ when $M$ is
diagonal). The matrix
\be
\Gamma = {1\over p!\sqrt{\det m}}\, \epsilon^{a_1\dots a_p}
\partial_{a_1}X^{I_1} 
\dots \partial_{a_p} X^{I_p} \gamma_0 \Gamma_{I_1\dots I_p}
\ee
will play an important role in what follows. It has the property that
\be
\Gamma^2 = (-1)^{(p-2)(p-5)/2} \,.
\ee
To verify this one notes first that, as for any product of Dirac matrices, 
\be
\Gamma^2 = \sum_k {1\over k!}C^{I_1\dots I_k}\Gamma_{I_1\dots I_k} \equiv \sum_k
C^{(k)} \cdot \Gamma_{(k)}\, ,
\ee
for some coefficient functions $C^{(k)}$; one then observes that $C^{(k)}$ must
vanish for $k\ne0$ because no antisymmetric tensor can be constructed from the
$(p-k)$ factors of the induced metric arising from the `contractions'
of Dirac matrices that must be made to get the k'th term. 
Evaluation of the zeroth term then yields the
result. We will restrict ourselves to the cases $p=2,5$, for which
\be\label{propxi}
\Gamma^2 \equiv 1\, .
\ee
We will also take $D=11$, so ${\cal S}$ is 10-dimensional. In this case the
2-brane and 5-brane have a natural interpretation as the M2-brane and M5-brane
of M-theory (with the tensor gauge field set to zero in the latter case). 

We may, and will, choose the $32\times 32$ D=11 Dirac matrices to be real. 
Let $\epsilon(X)$ be a real (commuting) time-independent $32$-component spinor
field on ${\cal M}$, normalized so that
\be\label{norm}
\epsilon^T \epsilon =1\, ,
\ee
and let $\Phi$ be the $p$-form on ${\cal S}$ defined by
\be
\Phi = {1\over p!} (\bar\epsilon \Gamma_{I_1\dots I_p}\epsilon)
\, dX^{I_1}\wedge \dots \wedge dX^{I_p}\, ,
\ee
where
\be
\bar\epsilon \equiv \epsilon^T\gamma_0\, .
\ee
We shall choose $\epsilon$ to be covariantly constant with respect to
a metric spin connection. In this case $\Phi$ is a closed form,
\be
d\Phi =0 \, .
\ee
This p-form $\Phi$ induces a p-form $\phi$ on $w$, given by
\be
\phi = vol\, (\epsilon^T \Gamma \epsilon)\, .
\ee
where
\be
vol\, \equiv d\sigma^1 \wedge \dots \wedge d\sigma^p\, \sqrt{\det m}
\ee
is the volume p-form on $w$ in the induced metric $m$. From the property
(\ref{propxi}) it follows that
\be\label{calineq}
\phi \le vol\, .
\ee
A closed p-form $\Phi$ with this property is called a (p-form) {\it
calibration} \cite{HL}. A p-surface in ${\cal S}$ for which this inequality is
everywhere saturated is said to be a {\it calibrated} surface, calibrated by
$\Phi$.

The significance of calibrations resides in their connection to minimal
p-surfaces \cite{HL}. Let $w$ be a calibrated surface  and let $U$ be an open
subset of $w$. Then, by hypothesis,
\be
vol(U) = \int_U \Phi \, .
\ee
Now let $V$ be any deformation of $U$ in ${\cal S}$ such that 
$U-V =\partial D$ where $D$ is some (p+1)-surface in ${\cal S}$.
Then
\be
\int_U \Phi = \int_V \Phi + \int_{D}d\Phi = \int_V\Phi\, .
\ee
where the second equality follows from the fact that $\Phi$ is a closed form.
Because $\Phi$ is a calibration we have
\be
\int_V\Phi \le vol(V)\, .
\ee
Putting everything together we deduce that
\be
vol(U)\le vol(V)\, ,
\ee
which shows that $w$ is a minimal surface. 

Given a p-surface $w$ we may evaluate on any of its tangent p-planes the
matrix $\Gamma$, and hence the p-form $\phi$ induced by the calibration p-form
$\Phi$. The p-surface will be calibrated by $\Phi$ if and only if
there exists a covariantly constant normalized spinor $\epsilon$ such that 
\be\label{gamcal}
\Gamma\epsilon =\pm \epsilon
\ee
for all p-planes tangent to $w$. Because of the identity (\ref{propxi}), this
equation is automatically satisfied for any {\sl given} tangent p-plane, the
solutions spanning a 16-dimensional subspace of spinor space. The
intersection of these spaces for all tangent p-planes is the solution space of
the equation (\ref{gamcal}), which therefore has dimension $\le 16$. For a
generic p-surface the dimension will vanish, so a generic p-surface is not
calibrated by $\Phi$, but special surfaces, which will necessarily be
minimal, may be. It follows that minimal surfaces can be found by seeking
solutions of (\ref{gamcal}).  These minimal surfaces have the feature that they
partially preserve the supersymmetry of the M-theory vacuum; this can be
understood either as a consequence of the `$\kappa$-symmetry' of super-brane
actions \cite{BDPS,BBS,kal} or directly from the spacetime supersymmetry 
algebra \cite{GPT}, but the details of this connection between supersymmetry
and calibrations will not be needed here.  

Examples are provided by a p-brane, with p=2 or p=5, in D=11 spacetimes of the
form (\ref{block}), for which ${\cal S}=\bE^p\times M$. In this case $M_{IJ}$
takes the form (\ref{Mform}) and the induced metric $m$ in the physical gauge is
given by (\ref{mform}). It then follows that
\bea
\det m &=& 1 + \bfn X^i\cdot \bfn X^j G_{ij} \nn
&& +\, {1\over2}(\bfn X^i\cdot
\bfn X^j) (\bfn X^k \cdot \bfn X^l)\left(G_{ij}G_{kl} -
G_{ik}G_{jl}\right) \nn
&& + \ \dots + \det\left(\bfn X^i\bfn X^j G_{ij}\right)\, .
\eea
We also have
\be\label{use}
\sqrt{\det m}\, \Gamma = \left(\sum_{k=0}^p 
{(-1)^{k(k+1)/2}\over k!}\gamma^{a_1\dots
a_k}\partial_{a_1}X^{i_1}\cdots\partial_{a_k}X^{i_k}\Gamma_{i_1\dots
i_k}\right) \Gamma_*
\ee
where $\Gamma_*$ is the constant matrix
\be
\Gamma_* \equiv \gamma_0\Gamma_{1\dots p}\, .
\ee

We are now in a position to find calibrated p-surfaces from the  
calibration condition (\ref{gamcal}). Consider first the brane theory vacuum;
in this case the calibration condition reduces (for one choice
of sign) to
\be\label{local}
\Gamma_*\epsilon = \epsilon\, .
\ee
Since $\Gamma_*^2=1$ and $\tr \Gamma_* =0$, this condition 
reduces by half the space 
spanned by covariantly constant spinors on ${\cal S}$. The calibrated p-surface
is a planar p-surface that fills the $\bE^p$ factor of ${\cal S}$. It is
calibrated by the p-form
\be
\Phi = dx^1\wedge \dots\wedge dx^p\, .
\ee
Since every p-surface is locally planar the condition (\ref{local}) must always
be satisfied, but for non-planar p-surfaces it will not be sufficient. To
determine the required additional conditions we can use (\ref{local}) in 
(\ref{gamcal}) to reduce the latter to 
\bea\label{calslag}
\sqrt{\det m}\, \epsilon &=& \bigg(1 - 
\gamma^a\partial_a X^i \Gamma_i - {1\over2}
\gamma^{ab}\partial_a X^i \partial_b X^j \Gamma_{ij} \nn
&& +\, {1\over6}\gamma^{abc}\partial_aX^i\partial_b X^j\partial_c X^k
\Gamma_{ijk} + \dots \bigg)\epsilon
\eea

The simplest non-trivial way to solve this condition is to suppose that each
power of $\partial X$ cancels separately. The cancellation of the linear term
requires 
\be\label{planar2}
\gamma^a \partial_a X^i \Gamma_i \epsilon =0\, .
\ee
Remarkably, this implies that all higher powers in $\partial X$ cancel
\cite{HL}. Here we shall verify this for $p=2$ \cite{BT}. 
Iteration of (\ref{planar2}) yields
\be\label{iterate}
-\gamma^{ab}\partial_a X^i\partial_b X^j \Gamma_{ij}\epsilon  = \bfn X^i \cdot
\bfn X^j G_{ij}\epsilon\, . 
\ee
Since $\epsilon$ is non-zero by hypothesis, the calibration condition is now
reduced to the condition
\be
\sqrt{\det (\bI + \tilde m)} = 1 + {1\over2}\tr \tilde m
\ee
where we have set
\be
\tilde m_{ab} = \partial_a X^i \partial_b X^j G_{ij}\, .
\ee
This condition is equivalent to 
\be
\tr \tilde m^2 = {1\over2}(\tr\tilde m)^2\, .
\ee
This is indeed a consequence of (\ref{planar2}) 
and can be proved by iteration of (\ref{iterate}) and use of the Dirac
matrix identity
\be
\Gamma_{IJ}\Gamma_{KL} = \Gamma_{IJKL} + 2 M_{L[I}\Gamma_{J]K} -
2 M_{K[I}\Gamma_{J]L} + 2M_{j[k}M_{L]I}\, .
\ee

We have just seen that we can find non-planar calibrated membranes in a
6-dimensional subspace $\bE^2\times M$ of the 10-dimensional space ${\cal S}$ 
by seeking fields $X^i(\sigma)$ for which (\ref{planar2}) admits non-zero
solutions for constant $\epsilon$. We are now going to make contact with the
results of the previous lecture by showing that solutions of the
Bogomol'nyi-type equation (\ref{CR}) are precisely the required configurations.
We will choose the top sign, for convenience, and rewrite this equation as
\be
\partial_2 X^i = \partial_1 X^j I_j{}^i\, .
\ee
Subsitution of this into (\ref{planar2}), followed by multiplication by
$\gamma^1$, yields
\be
\partial_1 X^i\left(\Gamma_i + 
\gamma^{12}\Gamma_j I_i{}^j\right)\epsilon  =0\, .
\ee
In coordinates for which $I$ takes a standard skew-diagonal form, with two
$2\times 2$ blocks, this becomes
\bea
&\bigg[e^1\left(\Gamma_1 + \gamma^{12}\Gamma_2\right) + 
e^2\left(\Gamma_2 - \gamma^{12}\Gamma_1\right) \nn
& \qquad \qquad +\, e^3\left(\Gamma_3 + \gamma^{12}\Gamma_4\right) +
e^4\left(\Gamma_4 - \gamma^{12}\Gamma_3\right) \bigg]\epsilon=0
\eea
where we have set $e^i=\partial_1 X^i$. This is equivalent to
\be
\sum_{k=1}^2 \left(\Gamma_{2k-1} e^{2k-1} + \Gamma_{2k}
e^{2k}\right)\left(1+\gamma^{12}
\Gamma_{2k-1}\Gamma_{2k}\right)\epsilon=0\, .
\ee
Thus, for each $k=1,2$, either $e^{2k-1}$ and $e^{2k}$ vanish, which is
equivalent to requiring one complex field to be constant, or
\be\label{formcon}
\left(1+ \gamma^{12}\Gamma_{2k-1}\Gamma_{2k}\right)\epsilon = 0\, .
\ee
Each such condition reduces the space of solutions of (\ref{planar2}) by 1/2.
The generic solution of (\ref{CR}) has all four scalar fields `active', and is
hence 1/4 supersymmetric. However, the generic solution does not have finite
energy. As we have seen, finite energy solutions correspond to membranes
wrapped on finite area holomorphic 2-cycles. Consider the two-centre model
discussed previously. The minimal energy membrane, wrapped on
the one finite-area holomorphic 2-cycle, has ${\bf X} = X{\bf n}$. It is
manifestly a configuration with two `active' scalars, which can be
replaced by the single complex scalar $Z= Xe^{i\phi}$. As a solution of the
supermembrane equations, this finite-energy lump solution is therefore 1/2
supersymmetric \cite{BT}.   

We have now seen how sigma model lump solutions of the membrane equations  
provide examples of calibrated surfaces. The calibration 2-form for this
special class of calibrated surfaces is called a {\sl K\"ahler} calibration,
for reasons that will now be explained. We begin by recalling that
\be
\Phi = {1\over 2}(\bar\epsilon \Gamma_{IJ}\epsilon)\, dX^IdX^J\, .
\ee
In the physical gauge, this becomes
\be
\Phi = {1\over2}(\bar\epsilon\gamma_{ab}\epsilon)\, d\sigma^a d\sigma^b + 
(\bar\epsilon\gamma_a\Gamma_i\epsilon)\, d\sigma^a dX^i 
+ (\bar\epsilon\Gamma_{ij}\epsilon)\, dX^i dX^j\, ,
\ee
the wedge product being understood here and in what follows. 
As a result of the constraint (\ref{local}), which is here equivalent to
$\gamma_{012}\epsilon =\epsilon$, 
and the normalization (\ref{norm}) of $\epsilon$, 
the first term equals $d\sigma^1d\sigma^2$.
Furthermore, when $X=X(\bfs)$ the term linear in $dX$ is
\be
(\bar\gamma_a\Gamma_I\epsilon)\, d\sigma^a d\sigma^b \partial_b X^i
= d\sigma^1 d\sigma^2 (\epsilon^T \gamma^b\Gamma_i
\gamma_{012}\epsilon)\,\partial_b X^i\, ,
\ee
but this vanishes on using the constraints  (\ref{local}) and (\ref{planar2}). 
We therefore drop the term linear in $dX$. In coordinates $(\varphi,{\bf X})$
for $M$ we are then left with
\be
\Phi = d\sigma^1 d\sigma^2 + 
(\bar\epsilon\Gamma_3\Gamma_a\epsilon) e^\phi e^a
+ {1\over2} (\bar\epsilon \Gamma_{ab}\epsilon) e^a e^b
\ee
where $(e^\varphi, e^a)$ are the frame 1-forms for which the 4-metric
takes the form
\be
ds^2 = (e^\varphi)^2 + \sum_{a=1}^3 (e^a)^2\, .
\ee

Referring to the M-theory array of the previous lecture, we see that
the 3-direction is the one with coordinate $\varphi$ and that we
should take $a=(3,4,5)$. The constraints (\ref{memconstraint} associated
with that array imply
\be
(\bar\epsilon\Gamma_{34}\epsilon)=(\bar\epsilon\Gamma_{56}\epsilon)=
1\, ,
\ee
and
\be
(\bar\epsilon\Gamma_{35}\epsilon) =
(\bar\epsilon\Gamma_{36}\epsilon) =
(\bar\epsilon\Gamma_{45}\epsilon) =
(\bar\epsilon\Gamma_{46}\epsilon) =0\, .
\ee
After relabelling $a=(4,5,6)\rightarrow (1,2,3)$ we then find that the
surviving terms in $\Phi$ are
\be
\Phi = d\sigma^1 d\sigma^2 +  e^\phi e^1 - e^2e^3\, .
\ee
That is 
\be
\Phi = d\sigma^1 d\sigma^2 + \Omega
\ee
where $\Omega = \bfO\cdot {\bf n}$ is the target space K\"ahler
2-form. Clearly, the calibration form $\Phi$ is a K\"ahler 2-form on 
the larger space $\bE^2\times M$; it is therefore called a K\"ahler
calibration.  

\section{Lecture 4: Beyond Field Theory}

So far, we have seen how field theory solitons are interpreted within brane
theory. Now we are going to see how brane theory allows additional static
`solitons' for which there is no field theory analogue. It will be useful to
begin by reviewing the {\it a priori} limitations due to Derrick's theorem.
Suppose that we have an energy functional of the form
\be
E[X]= \sum_k E_k[X]\, ,
\ee
where the functionals $E_k$ are p-dimensional integrals with
integrands that are homogeneous of degree $k$ in derivatives of a set of
scalar fields $X(\sigma)$. To any given field configuration ${\overline
X}(\sigma)$ corresponds a value ${\overline E}_k$ of $E_k$ and hence an energy
${\overline E}$. Now ${\overline E}_k\rightarrow \lambda^{k-p}{\overline E}_k$
under a uniform scaling $\sigma \rightarrow \lambda \sigma$ of the coordinates
$\sigma$, so a necessary condition for ${\overline X}(\sigma)$ to minimise $E$
is that
\be\label{derrick}
\sum_k (k-p){\overline E}_k =0\, ,
\ee
since there otherwise exists a $\lambda$ for which the field
configuration $X(\sigma)={\overline X}(\lambda \sigma)$ has lower energy. If
$E_k$ is non-negative for all $k$ in the sum then for any finite sum there will
be a $p$ for which $(k-p)$ is always negative and (\ref{derrick}) cannot be
satisfied unless $E_k=0$ for all $k\ne p$. For conventional scalar field 
theories one has $k=0,2$, and since ${\overline E}_2$ vanishes only in the
vacuum there can be no static solitons for $p>2$. This is Derrick's theorem. A
corollary is that when $p=2$ we must have ${\overline E}_0=0$, which can
normally be satisfied only if the scalar potential vanishes.

For a brane theory the energy functional is non-polynomial in
derivatives and Derrick's theorem no longer applies. Of course, solutions that
are `Derrick-forbidden' must involve a cancellation of terms of different
scaling weight and cannot be solutions of {\sl first-order} equations
of Bogomol'nyi-type; the relevant equations are necessarily non-linear in
derivatives. The simplest example is provided by a 3-brane in $\bE^6$. The
$\bE^6$ coordinates are $(X^a,X^i)$ with $a=(1,2,3)$ and $i=(4,5,6)$. In a
physical gauge we have $X^a=\sigma^a$, where $\sigma^a$ are the worldspace
coordinates, and $X^i(\bfs)$ are the physical fields. 
Let us define the 3-vector
\be
{\bf X}=(X^4,X^5,X^6) 
\ee
so that, for example, $\bfn \cdot {\bf X} \equiv \tr (\partial X)$. In the
physical gauge,
\be
({\cal E}+1)^2  = \det \left(\bI + (\partial X) (\partial X)^T\right)
\ee
Now, we use the identity\footnote{This is the $3\times 3$ case of an identity
given by Harvey and Lawson for the $n\times n$ case \cite{HL}. I thank Jerome
Gauntlett for pointing this out and for helping to transcribe the $3\times
3$ result to the notation used here.}
\bea\label{HL2}
\det \left(\bI + (\partial X) (\partial X)^T\right) &\equiv & \left[1 - \star
\psi\right]^2 + \left(\bfn \cdot {\bf X} -
\det \partial X \right)^2 \nn 
&+& |\bfn \times {\bf X}|^2 + \sum_{i=4}^6 \left(\bfn X^i \cdot \bfn \times
{\bf X}\right)^2
\eea
where $\star\, \psi$ is the worldspace dual of the closed worldspace 3-form
\be\label{psi}
\psi = {1\over2} d\bfs \cdot d{\bf X} \times d{\bf X}\, ,
\ee
the wedge product of forms being implicit. Given that $\star\psi$ is
negative\footnote{This assumption is necessary because, in contrast to
the analogous identity for K\"ahler calibrations we are {\sl not} free
to adjust the signs in the identity (\ref{HL2}). It is possible to
find configurations for which $\star\psi$ is positive, and even such
that $(1-\star\psi)$ is negative, but the simplest examples are such
that $|{\bf X}|$ does not vanish as $|\bfs|\rightarrow
\infty$. Presumably, this condition guarantees that 
$\star\psi \le 0$.} we may deduce the bound
\be
{\cal E}\ge |\psi|
\ee
with equality when 
\be\label{slagg}
\bfn \times {\bf X} ={\bf 0} \, ,\qquad
\bfn \cdot {\bf X} =  \det \partial X\, .
\ee
These equations (\ref{slagg}) describe a {\it special Lagrangian} (SLAG) 
3-surface in $\bE^6$. Note
that these conditions combined with (\ref{HL2}) imply that
\be\label{sqrtm}
\sqrt{\det m} = 1 -  \star \psi\, .
\ee 
The curl-free condition is equivalent to
\be
d{\bfs}\cdot d{\bf X}=0\, .
\ee
The left hand side is a symplectic 2-form on $\bE^6$ (the wedge product of
forms again being implicit). A lagrangian submanifold is a 3-surface on which
this form vanishes. Since ${\bf X}$ is curl free we have, locally,
\be\label{principalf}
{\bf X} = \bfn S
\ee
for some scalar function $S(\bfs)$ of the three worldspace coordinates.
Any such function provides a local description of a Lagrangian 3-surface.
The additional `special' condition is needed for it to be minimal. In terms of
$S$, this condition is 
\be\label{hess}
\nabla^2 S =  \det {\rm Hess} S
\ee
where the {\sl Hessian} of $S$ is the matrix of second partial derivatives of
$S$. 

We are now going to see how these equations can be understood via the theory of
calibrations. For the K\"ahler calibrations considered previously, the
calibration condition (\ref{calslag}) was satisfied order by order in an 
expansion in powers of $\partial X$. This was to be expected from the fact that
the `BPS' condition was homogeneous in derivatives. Now we should expect to 
satisfy (\ref{calslag}) by a cancellation between different powers of $\partial
X$. Special Lagrangian 3-surfaces in $\bE^6$ have an M-theory
interpretation in terms of three M5-branes intersecting according to
the array \cite{GLW}
$$
\begin{array}{lcccccccccccc}
M5: & 1 & 2 & 3 & - & - & - & | & 7 & 8 & - & - & - \nn
M5: & - & - & 3 & 4 & 5 & - & | & 7 & 8 & - & - & - \nn
M5: & - & 2 & - & 4 & - & 6 & | & 7 & 8 & - & - & -
\end{array}
$$ 
Omitting the two common worldspace directions, and the last two
transverse directions, neither of which plays a role, we have effectively three
3-branes in $\bE^6$. We can read off from the array the conditions imposed on
the spinor $\epsilon$ by these three branes, up to a choice of signs. For
example
\be
\Gamma_{012378}\epsilon = \epsilon\, ,\qquad
\Gamma_{034578}\epsilon = -\epsilon\, ,\qquad
\Gamma_{024678}\epsilon= \epsilon\, . 
\ee
Each product of Dirac matrices on the left hand side of these equations has
eigenvalues $\pm1$, and the corresponding constraint projects out one of these
eigenspaces according to the sign chosen; the signs here have been chosen for
convenience. Note that these constraints imply
\be\label{gamconds}
\Gamma^{1245}\epsilon = \epsilon\, ,\qquad
\Gamma^{1346}\epsilon = \epsilon\, ,\qquad
\Gamma^{2356}\epsilon = \epsilon\, .
\ee

The above discussion assumes that the only constraints are those associated
with the three tangent planes indicated in the array. This is obviously the
case if the configuration represented by the array is a singular
orthogonal intersection of three planar M5-branes, but it may be possible to
smooth the intersection in such a way that no further constraints 
arise, in which case the whole configuration can be interpreted as 
a single M5-brane asymptotic to the three M5-branes of the array. Our
aim is to find the equations that govern such smooth intersections. 
We may choose the first of the asymptotic planar M5-branes
as the M5-brane vacuum, interpreting the rest as a `solitonic'
deformation about this vacuum. Note that the first constraint is then the
vacuum constraint $\Gamma_*\epsilon =\epsilon$. Imposing this condition, and 
taking
\be
\Gamma_a \rightarrow \gamma_a ,\qquad (a=1,2,3)
\ee
to accord with our earlier notation, we again arrive at (\ref{calslag}), 
but we will no longer assume that the terms linear and cubic in
$\partial X$ must vanish separately. Instead we allow for the possibility that
that they may conspire to cancel; noting that 
\be
{1\over6}\gamma^{abc}\partial_aX^i\partial_bX^j\partial_cX^k\Gamma_{ijk} 
= \gamma^{123}\Gamma_{456}\det(\partial_a X^i)
\ee
and that $\gamma^{23}\Gamma_{56}\epsilon=\epsilon$, this cancellation requires
\be\label{slag}
\gamma^a\partial_a X^i \Gamma_i\epsilon = \det(\partial X)\,
\gamma^1\Gamma_4\epsilon\, .
\ee
We now observe that  (\ref{gamconds}) implies 
\be
\gamma^1\Gamma_4 \epsilon = \gamma^2\Gamma_5 \epsilon = 
\gamma^3\Gamma_6 \epsilon 
\ee
and 
\be
\gamma^3\Gamma_5 \epsilon = - \gamma^2\Gamma_6\epsilon\, ,\qquad
\gamma^1\Gamma_6 \epsilon = - \gamma^3\Gamma_4\epsilon\, ,\qquad
\gamma^2\Gamma_4 \epsilon = - \gamma^1\Gamma_5\epsilon\, .
\ee
These constraints imply, in turn, that
\be
\gamma^a\partial_a X^i \Gamma_i\, \epsilon= \left[( \bfn \cdot {\bf
X})\gamma^1\Gamma_4  + (\bfn \times {\bf
X})\cdot {\bf G}\right] \epsilon\, ,
\ee
where we have set
\be
{\bf G} = (\gamma^2\Gamma_6,\gamma^3\Gamma_4,\gamma^1\Gamma_5).
\ee
Putting all this together we see that (\ref{slag}) is satisfied 
if and only if ${\bf X}(\bfs)$ satisfies (\ref{slagg}). Thus 
the special Lagrangian equations are {\sl necessary} for a smooth
calibrated intersection. We next show that they are also 
{\sl sufficient}.   

Using (\ref{sqrtm}) the calibration condition becomes 
\be\label{calred}
\left[1 -\star\psi\right] \epsilon =
\left(1- {1\over2}\gamma^{ab}\partial_aX^i\partial_bX^j\Gamma_{ij}
\right)\epsilon =\, ,
\ee
since the terms linear and cubic in $\partial X$ on the right hand
side have cancelled. We will see that this condition is identically
satisfied, without any further conditions imposed on $\epsilon$. 
Firstly, interation of (\ref{slag}), and 
further use of (\ref{gamconds}), yields
\bea
\left(\gamma^a\partial_a X^i\Gamma_i\right)^2 \epsilon &=&
-(\det \partial X)^2 \epsilon +  2 \left[(\bfn \times {\bf X})_3 \Gamma_{45} -
(\bfn \times {\bf X})_2 
\Gamma_{46}\right]\epsilon\, \nn
&=& (\tr \partial X)^2 \epsilon \, ,
\eea
where (\ref{slagg}) has been used to arrive at the second line.
Multiplying out the Dirac matrices on the right hand side, and using
(\ref{slagg}) again, we find that
\be
\gamma^{ab}\partial_a X^i\partial_b X^j \Gamma_{ij} = (\tr \partial X)^2 - \tr
(\partial X)^2 \, .
\ee
The calibration condition (\ref{calred}) is thus equivalent to
\be\label{sat}
\star\psi  = {1\over2}\tr (\partial X)^2 -{1\over2}(\tr \partial X)^2
\ee
Howe
but this is identically satisfied as a consequence of 
the special Lagrangian conditions (\ref{slagg}).

Finally, we turn to the relation between the 3-form $\psi$ of (\ref{psi}) and
the calibration 3-form $\Phi$. Recall that
\be
\Phi = {1\over 6}(\bar\epsilon \Gamma_{IJK}\epsilon)\, dX^I dX^J dX^K
\ee
for the case at hand. On going to the physical gauge we can expand the right
hand side in powers of $\partial_a X^i$. Because the linear and cubic terms
cancel on the calibrated surface we may drop these terms. What is left is the
zeroth term and the quadratic term, and these are
\be\label{finalphi}
\Phi = {1\over 6}d\bfs \cdot d\bfs \times d\bfs - \psi\, . 
\ee  
We can rewrite this as
\be
\Phi = {\cal R}{\rm e}\left[ d{\bf Z}\cdot d{\bf Z} \times d{\bf Z}\right]\, ,
\ee
where ${\bf Z} = \bfs + i {\bf X}$ is a set of complex 
coordinates for $\bC^3$. 
This illustrates a general feature of special Lagrangian calibrations. A
special Lagrangian p-surface in $\bC^p$ is calibrated by the real part of a 
holomorphic p-form. For $p=2$ this is the real part of a holomorphic 2-form,
which we can identify as the K\"ahler 2-form $\Omega$ of the previous
lecture. 

Of course, to find finite energy SLAG solitons of the type disussed
we would need to choose a background with a holomorphic 3-cycle of
finite 3-volume and admitting a covariantly constant holomorphic
3-form, but this I leave until such time as I have understood it better.
In the meantime, the reader is invited to consult \cite{lust} for some
interesting applications of SLAG calibrations. I should not leave the
impression that K\"ahler and SLAG calibrations are the only cases. There are
also some `exceptional' calibrations. These also have a realization in terms of
intersecting M5-branes and I refer to \cite{phrem} for a recent review of some
applications.

\section{Epilogue: the brane world}

These lectures have argued that field theory can be understood as a limit of 
a more encompassing `brane theory'. Brane theory identifies certain field theory
solitons with minimal surfaces (in spaces of reduced holonomy) associated to
simple calibrations, but it goes beyond field theory in allowing other types of
minimal surface, associated with more complicated calibrations. Only scalar field
theories were considered here but a similar case can be made for gauge theories
via D-branes and generalizations of calibration theory to include worldvolume
gauge fields
\cite{GLW2,moore}.  Gravitational field theories, on the other hand, do not
have an analogous brane theory interpretation because gravitons (and
superpartners) propagate in the `bulk' and not on branes. This explains the
universality of gravitational interactions\footnote{ Branes may still describe gravity
via an equivalence to field theory, as in the M(atrix) model and the adS/CFT
correspondence, but the variety of these equivalences corresponds to varieties
of supergravity theories and not to varieties of graviton.}; while there can be
many branes, and many types of brane, there is only one `bulk'.

Thus, `brane theory' is naturally non-gravitational. Of course, there could
still be an effective gravity at sufficiently low energy if the bulk is 
compact. However, in this case we are dealing with Kaluza-Klein theory rather
than brane theory. One cannot really consider the lower-dimensional spacetime
as a brane in this case because this brane is not localized in the extra
dimension. This is a necessary feature in the quantum theory since the
uncertainty principle guarantees complete delocalization in a compact space at
zero momentum. Until recently it used to be thought that any decompactification
of the bulk would cause a loss of localization of the graviton on the
lower-dimensional space, but Randall and Sundrum have shown that this need not
be the case; in particular, gravity is localized on a horospherical boundary in
anti-de Sitter space. This boundary is called a `Brane World'. The
brane interpretation is problematic, however, because gravity couples to the
energy-momentum-stress tensor of matter, and `brane matter' does not have a
conventional stress tensor. 

The symmetric stress tensor $T^{\mu\nu}$ for a Minkowski spacetime field theory
is essentially the set of four Noether currents associated with translational
invariance. One way to find the Noether currents is to note that the variation
of the action under an infinitesimal but {\sl non-uniform} translation with
parameters $\alpha$ must be of the form
\be\label{varS}
\delta_\alpha S = \int j \wedge d \alpha\, .
\ee
The left hand side vanishes (off-shell) when $d\alpha=0$ but it must vanish
on-shell even when $d\alpha \ne0$, so the coefficient form $j$ must be closed
on-shell. Its dual vector density is therefore conserved, on shell, and can be
identified as the Noether current. However, this prescription fails to define a
translation Noether current for a brane theory because the action is not only
invariant under uniform translations but also non-uniform translations, so the
variation (\ref{varS}) vanishes identically. 

This problem can be circumvented by fixing the worldvolume reparameterizations
before applying the Noether prescription. In the physical gauge, $X^\mu =
\sigma^\mu$, we have
$g=\eta +\tilde g$, where
\be
\tilde g_{\mu\nu} = \partial_\mu X^i\partial_\nu X^j G_{ij}(X)
\ee
and $X^i$ are the physical worldvolume fields. The Lagrangian is now
\be
{\cal L} = -\sqrt{-\det (\eta + \tilde g)}\, .
\ee
The stress tensor is given by
\cite{LW}
\be
\sqrt{-\det\eta}\, T^{\mu\nu} = \sqrt{-\det g}\, g^{\mu\nu}
\ee
and $\partial_\mu T^{\mu\nu}=0$ in cartesian coordinates, on shell, because the
$X^\mu$ field equation in physical gauge is
\be
\partial_\mu \left(\sqrt{-\det g}\,g^{\mu\nu}\right)=0
\ee
This looks non-covariant but that is to be expected after gauge-fixing. 

If this stress tensor is used to couple the brane to worldvolume gravity one
finds the Lagrangian
\be\label{laggrav}
{\cal L}= -\sqrt{-\det(g + h)}\, ,
\ee
where $h \equiv \gamma - \eta$ is the deviation of the {\sl independent}
worldvolume metric $\gamma$ from the Minkowski metric. Since $g=\eta +\tilde g$
we can write this as
\be
{\cal L}= -\sqrt{-\det(\gamma + \tilde g)}\, .
\ee
An expansion in powers of $\tilde g$ now yields the sum of a cosmological term
and a more-or-less standard action for scalar fields $X^i$ coupled to gravity
via the metric $\gamma$. From this interpretation one can see that we have now
{\sl regained} reparameterization invariance. But we have paid a price: since 
one cannot fix a gauge twice, the interpretation as a gauge-fixed brane action
has been lost. If one attempts to recover this interpretation by returning to
(\ref{laggrav}) and taking $g$ to be the induced metric {\sl prior} to choice
of the physical gauge then one has a coupling to gravity that is explicitly
background dependent since it depends on the perturbation $h$ and not the full
metric $\gamma$. 

In the case of an adS background, the problem can be phrased in a different
way. The action is invariant under all isometries of the background, but from
the point of view of the brane these are symmetries of a non-linearly realized
conformal invariance \cite{malda,Kallosh}. This observation applies, in
particular, to a brane for which the worldvolume is a horosphere near the adS
boundary. Such a brane provides a realization of the Randall-Sundrum mechanism
by which gravity is induced on the brane \cite{RS}, but the coupling to gravity
on the brane will now break the non-linearly realized conformal invariance. On
the other hand, this brane is supposed to be equivalent to a CFT on the adS
boundary with a UV cut-off \cite{gubser}. But a cut-off also breaks
conformal symmetry. From this perspective it is no surprise that gravity on the
brane breaks the non-linearly realized conformal symmetry of a brane action.

\vskip 0.5cm
\noindent
{\bf Acknowledgements}: I am grateful to Jerome Gauntlett for several very
useful discussions on the content of these lectures.


\bigskip

\begin{thebibliography}{99} 



\bibitem{Marija}
J.P. Gauntlett, C. Koehl, D. Mateos, P.K. Townsend and M. Zamaklar, {\sl Finite
energy Dirac-Born-Infeld monopoles and string junctions}, 
Phys. Rev. {\bf D60} (1999) 045004. 

\bibitem{BT}
E. Bergshoeff and P.K. Townsend, {\sl Solitons on the Supermembrane}, JHEP
9905:021 (1999).

\bibitem{albert}
A.S. Schwarz, {\sl Supergravity and field-space democracy}, Nucl. Phys. {\bf B
171}, (1980) 154; A.V. Gayduk, V.N. Romanov and A.S. Schwarz, {\sl
Supergravity and field-space democracy}, Commun. Math. Phys. {\bf 79} (1981)
507. 

\bibitem{GGT}
J.P. Gauntlett, J. Gomis and P.K. Townsend, {\sl BPS bounds for worldvolume
branes}, JHEP 9801:003 (1998).

\bibitem{HL}
R. Harvey and H.B. Lawson, {\sl Calibrated geometries}, Acta. Math. {\bf 148}
(1982) 47.

\bibitem{GPT}
J. Gutowski, G. Papadopoulos and P.K. Townsend, {\sl Supersymmetry and
generalized calibrations}, Phys. Rev. {\bf D60} (1999) 106006.

\bibitem{BBS}
K. Becker, M. Becker and A. Strominger, {\sl Fivebranes, membranes and
non-perturbative string theory}, Nucl. Phys. {\bf B456} (1995) 130.

\bibitem{oog}
K. Becker, M. Becker, D.R. Morrison, H. Ooguri and Z. Yin, {\sl Supersymmetric
cycles in exceptional holonomy manifolds and Calabi-Yau 4-folds}, Nucl. Phys.
{\bf B480} (1996) 225.

\bibitem{GibPap}
G.W. Gibbons and G. Papadopoulos, {\sl Calibrations and intersecting branes},
Commun. Math. Phys. {\bf 202} (1999) 571. 

\bibitem{GLW}
J.P. Gauntlett, N.D. Lambert and P.C. West, {\sl Branes and calibrated
geometries}, Commun. Math. Phys. {\bf 202} (1999) 593.

\bibitem{QMW}
B.S. Acharya, J.M. Figueroa-O'Farrill and B. Spence, {\sl Branes at angles and
calibrations}, JHEP {\bf 04:012} (1998).

\bibitem{RS}
L. Randall and R. Sundrum, {\sl An alternative to compactification}, Phys. Rev.
Lett. {\bf 83} (1999) 4690.

\bibitem{azco}
C. Chryssomalakos, J.A. de Azc{\'a}rraga, J.M. Izquierdo and 
J.C. P{\'e}rez-Bueno, {\sl The geometry of branes and extended
superspaces}, Nucl. Phys. {\bf 567B} (2000) 293.

\bibitem{BSTanii}
E. Bergshoeff, E. Sezgin, Y. Tanii, {\sl Hamiltonian formulation of the
supermembrane}, Nucl. Phys. {\bf B298} (1988) 187;\\ 
J.A. de Azc{\' a}rraga, J.M. Izquierdo and P.K. Townsend, {\sl Classical
anomalies of supersymmetric extended objects}, Phys. Lett. 
{\bf 267B} (1991) 366.
 

\bibitem{HP}
J. Hughes and J. Polchinski, {\sl Partially broken global supersymmetry and the
superstring}, Nucl. Phys. {\bf B278} (1986) 147;\\ J.P. Gauntlett, {\sl Current
algebra of the D=3 superstring and partial breaking of supersymmetry}, Phys.
Lett. {\bf 228B} (1989) 188;\\ A. Ach{\' u}carro, 
J. Gauntlett, K. Itoh and P.K.
Townsend, {\sl Worldvolume supersymmetry from spacetime supersymmetry of the
four dimensional supermembrane}, Nucl. Phys. {\bf B314} (1989) 129.

\bibitem{AGIT}
J. A. de Azc{\' a}rraga, J.P. Gauntlett, J.M. Izquierdo 
and P.K. Townsend, {\sl
Topological extensions of the supersymmetry algebra for 
extended objects}, Phys.
Rev. Lett. {\bf 63} (1989) 2443.

\bibitem{zumino}
B. Zumino, {\sl Supersymmetry and K\"ahler manifolds}, Phys. Lett. {\bf 87B}
(1979) 203.

\bibitem{AGF}
L. Alvarez-Gaum{\'e} and D.Z. Freedman, {\sl Ricci flat Kahler manifolds and
supersymmetry}, Phys. Lett. {\bf 94B} (1980) 171. 

\bibitem{per}
A.M. Perelomov, {\sl Instantons and K\"ahler manifolds}, 
Commun. Math. Phys. {\bf 63} (1978) 237;\\
R.S. Ward, {\sl Slowly moving lumps in the $CP^1$ model in
(2+1)-dimensions}, Phys. Lett. {\bf 158B} (1985) 424;\\
P.J. Ruback, {\sl Sigma model solitons and their moduli space metrics}, Commun.
Math. Phys. {\bf 116} (1988) 645.

\bibitem{gibrub}
G.W. Gibbons and P.J. Ruback, {\sl The hidden symmetries of multi-centre
metrics}, Commun. Math. Phys. {\bf 115} (1988) 267.

\bibitem{bogomol}
E.B. Bogomol'nyi, {\sl The stability of classical solutions},
Sov. J. Nucl. Phys. {\bf 24} (1976) 449.

\bibitem{CMG} 
C. Callan and J. Maldacena, {\sl Brane dynamics from the 
Born-Infeld action}, Nucl. Phys. {\bf B513} (1998) 198; \\
G.W. Gibbons, {\sl Born-Infeld particles and Dirichlet p-branes}, 
Nucl. Phys. {\bf B514} (1998) 603.

\bibitem{BSTown}
E. Bergshoeff, E. Sezgin and P.K. Townsend, {\sl Supermembranes and 11
dimensional supergravity}, Phys. Lett. {\bf 189B} (1987) 75; {\sl Properties of
the eleven-dimensional supermembrane theory}, 
Ann. Phys. (N.Y.) {\bf 185} (1988) 330.

\bibitem{han}
S.K. Han and I.G. Koh, {\sl N=4 remaining supersymmetry in a Kaluza-Klein
monopole background in D=11 supergravity}, Phys. Rev. {\bf D31} (1985) 2503;\\
P.K. Townsend, {\sl The eleven-dimensional supermembrane revisited}, 
Phys. Lett.
{\bf 350B}, (1995) 184.


\bibitem{carg}
P.K. Townsend, {\sl M-theory from its superalgebra}, in 
{\it Strings, Branes and
Dualities}, (Carg{\`e}se lectures 1997), eds. L. Baulieu, P. Di
Francesco, M. Douglas, V. Kazakov, M. Picco and P. Windey 
(Kluwer 1999) pp. 141-177; hep-th/9712004.

\bibitem{BDPS}
E. Bergshoeff, M.J. Duff, C.N. Pope and E. Sezgin, {\sl Supersymmetric
supermembrane vacua and singletons}, Phys. Lett. {\bf 199B} (1987) 69.

\bibitem{kal}
E. Bergshoeff, R. Kallosh, T. Ort{\'\i}n and 
G. Papadopoulos, {\sl Kappa-symmetry,
supersymmetry and intersecting branes}, Nucl. Phys. {\bf B502} (1997)
149.

\bibitem{lust}
A. Karch, D. L\"ust and A. Miemiec, {\sl N=1 supersymmetric gauge
theories and supersymmetric 3-cycles}, Nucl. Phys. {\bf B553} (1999)
483.

\bibitem{phrem}
P.K. Townsend, {\sl PhreMology: calibrating M-branes},
Class. Quantum Grav. {\bf 17} (2000) 1267. 

 \bibitem{GLW2}
J.P. Gauntlett, N.D. Lambert and P.C. West, {\sl Supersymmetric fivebrane
solitons}, Adv. Theor. Math. Phys. {\bf 3} (1999) 106006;\\
J.P. Gauntlett, {\sl Membranes on fivebranes}, hep-th/9906162

\bibitem{moore}
M. Marino, R. Minasian, G. Moore, A. Strominger, {\sl Non-linear instantons
from supersymmetric p-branes}, JHEP 0001:005 (2000).

\bibitem{LW}
O. Barwell, N.D. Lambert and P.C. West, {\sl On the energy momentum tensor of
the M-theory fivebrane}, Phys. Lett. {\bf B459} (1999) 125. 

\bibitem{malda}
J. Maldacena, {\sl The large N limit of superconformal field theories
and supergravity}, Adv. Theor. Math. Phys. {\bf 2} (1998) 231.

\bibitem{Kallosh}
P. Claus, R. Kallosh, J. Kumar, P.K. 
Townsend and A. Van Proeyen, {\sl Conformal
theory of M2, D3, M5 and `D1+D5' branes}, JHEP 9806:004 (1998).

\bibitem{gubser}
S. Gubser, {\sl AdS/CFT and gravity}, hep-th/9912001;\\
C.S. Chang, P.L. Paul and H. Verlinde, {\sl A note on warped string
compactification}, hep-th/0003236.


\end{thebibliography}
\end{document}